\documentclass[twocolumn]{aastex7}

\graphicspath{{./}{figures/}}

\usepackage{amsmath}
\usepackage{bm}
\usepackage{color}
\usepackage{hyperref}
\usepackage{subfig}

\newcommand{\frb}{FRB~20240114A}

\begin{document}

\title{The CHIME/FRB Discovery of the Extremely Active Fast Radio Burst Source FRB~20240114A}

\author[0000-0002-6823-2073]{Kaitlyn Shin}
  \affiliation{MIT Kavli Institute for Astrophysics and Space Research, Massachusetts Institute of Technology, 77 Massachusetts Ave, Cambridge, MA 02139, USA}
  \affiliation{Department of Physics, Massachusetts Institute of Technology, 77 Massachusetts Ave, Cambridge, MA 02139, USA}
  \email[show]{kshin@mit.edu}
  \correspondingauthor{Kaitlyn Shin}
\author[0000-0002-8376-1563]{Alice Curtin}
  \affiliation{Department of Physics, McGill University, 3600 rue University, Montr\'eal, QC H3A 2T8, Canada}
  \affiliation{Trottier Space Institute at McGill University, 3550 rue University, Montr\'eal, QC H3A 2A7, Canada}
  \email{alice.curtin@mail.mcgill.ca}
\author[0009-0006-1258-4228]{Maxwell Fine}
  \affiliation{Anton Pannekoek Institute for Astronomy, University of Amsterdam, Science Park 904, 1098 XH Amsterdam, The Netherlands}
  \email{max.fine@student.uva.nl}
\author[0000-0002-8897-1973]{Ayush Pandhi}
  \affiliation{David A.~Dunlap Department of Astronomy \& Astrophysics, University of Toronto, 50 St.~George Street, Toronto, ON M5S 3H4, Canada}
  \affiliation{Dunlap Institute for Astronomy \& Astrophysics, University of Toronto, 50 St.~George Street, Toronto, ON M5S 3H4, Canada}
  \email{ayush.pandhi@mail.utoronto.ca}
\author[0000-0002-3980-815X]{Shion Andrew}
  \affiliation{MIT Kavli Institute for Astrophysics and Space Research, Massachusetts Institute of Technology, 77 Massachusetts Ave, Cambridge, MA 02139, USA}
  \affiliation{Department of Physics, Massachusetts Institute of Technology, 77 Massachusetts Ave, Cambridge, MA 02139, USA}
  \email{shiona@mit.edu}
\author[0000-0002-3615-3514]{Mohit Bhardwaj}
  \affiliation{McWilliams Center for Cosmology, Department of Physics, Carnegie Mellon University, Pittsburgh, PA 15213, USA}
  \email{mohitb@andrew.cmu.edu}
\author[0000-0002-2878-1502]{Shami Chatterjee}
  \affiliation{Cornell Center for Astrophysics and Planetary Science, Cornell University, Ithaca, NY 14853, USA}
  \email{shami@astro.cornell.edu}
\author[0000-0001-6422-8125]{Amanda M. Cook}
  \affiliation{Department of Physics, McGill University, 3600 rue University, Montr\'eal, QC H3A 2T8, Canada}
  \affiliation{Trottier Space Institute at McGill University, 3550 rue University, Montr\'eal, QC H3A 2A7, Canada}
  \affiliation{Anton Pannekoek Institute for Astronomy, University of Amsterdam, Science Park 904, 1098 XH Amsterdam, The Netherlands}
  \email{amanda.cook@mcgill.ca}
\author[0000-0001-8384-5049]{Emmanuel Fonseca}
  \affiliation{Department of Physics and Astronomy, West Virginia University, PO Box 6315, Morgantown, WV 26506, USA }
  \affiliation{Center for Gravitational Waves and Cosmology, West Virginia University, Chestnut Ridge Research Building, Morgantown, WV 26505, USA}
  \email{emmanuel.fonseca@mail.wvu.edu}
\author[0000-0002-3382-9558]{B.~M.~Gaensler}
  \affiliation{Department of Astronomy and Astrophysics, University of California, Santa Cruz, 1156 High Street, Santa Cruz, CA 95064, USA}
  \affiliation{David A.~Dunlap Department of Astronomy \& Astrophysics, University of Toronto, 50 St.~George Street, Toronto, ON M5S 3H4, Canada}
  \affiliation{Dunlap Institute for Astronomy \& Astrophysics, University of Toronto, 50 St.~George Street, Toronto, ON M5S 3H4, Canada}
  \email{gaensler@ucsc.edu}
\author[0000-0003-2317-1446]{Jason Hessels}
  \affiliation{Department of Physics, McGill University, 3600 rue University, Montr\'eal, QC H3A 2T8, Canada}
  \affiliation{Trottier Space Institute at McGill University, 3550 rue University, Montr\'eal, QC H3A 2A7, Canada}
  \affiliation{Anton Pannekoek Institute for Astronomy, University of Amsterdam, Science Park 904, 1098 XH Amsterdam, The Netherlands}
  \affiliation{ASTRON, Netherlands Institute for Radio Astronomy, Oude Hoogeveensedijk 4, 7991 PD Dwingeloo, The Netherlands}
  \email{jason.hessels@mcgill.ca}
\author[0009-0009-0938-1595]{Naman Jain}
  \affiliation{Trottier Space Institute at McGill University, 3550 rue University, Montr\'eal, QC H3A 2A7, Canada}
  \email{naman.jain@mail.mcgill.ca}
\author[0000-0001-9345-0307]{Victoria M.~Kaspi}
  \affiliation{Department of Physics, McGill University, 3600 rue University, Montr\'eal, QC H3A 2T8, Canada}
  \affiliation{Trottier Space Institute at McGill University, 3550 rue University, Montr\'eal, QC H3A 2A7, Canada}
  \email{victoria.kaspi@mcgill.ca}
\author[0009-0008-6166-1095]{Bikash Kharel}
  \affiliation{Department of Physics and Astronomy, West Virginia University, PO Box 6315, Morgantown, WV 26506, USA }
  \affiliation{Center for Gravitational Waves and Cosmology, West Virginia University, Chestnut Ridge Research Building, Morgantown, WV 26505, USA}
  \email{bk0055@mix.wvu.edu}
\author[0000-0003-2116-3573]{Adam E. Lanman}
  \affiliation{MIT Kavli Institute for Astrophysics and Space Research, Massachusetts Institute of Technology, 77 Massachusetts Ave, Cambridge, MA 02139, USA}
  \affiliation{Department of Physics, Massachusetts Institute of Technology, 77 Massachusetts Ave, Cambridge, MA 02139, USA}
  \email{alanman@mit.edu}
\author[0000-0002-5857-4264]{Mattias Lazda}
  \affiliation{David A.~Dunlap Department of Astronomy \& Astrophysics, University of Toronto, 50 St.~George Street, Toronto, ON M5S 3H4, Canada}
  \affiliation{Dunlap Institute for Astronomy \& Astrophysics, University of Toronto, 50 St.~George Street, Toronto, ON M5S 3H4, Canada}
  \email{mattias.lazda@mail.utoronto.ca}
\author[0000-0002-4209-7408]{Calvin Leung}
  \affiliation{Miller Institute for Basic Research, University of California, Berkeley, CA 94720, USA}
  \affiliation{Department of Astronomy, University of California, Berkeley, CA 94720, USA}
  \email{calvin_leung@berkeley.edu}
\author[0000-0002-7164-9507]{Robert Main}
  \affiliation{Department of Physics, McGill University, 3600 rue University, Montr\'eal, QC H3A 2T8, Canada}
  \affiliation{Trottier Space Institute at McGill University, 3550 rue University, Montr\'eal, QC H3A 2A7, Canada}
  \email{robert.main@mcgill.ca}
\author[0000-0002-4279-6946]{Kiyoshi W. Masui}
  \affiliation{MIT Kavli Institute for Astrophysics and Space Research, Massachusetts Institute of Technology, 77 Massachusetts Ave, Cambridge, MA 02139, USA}
  \affiliation{Department of Physics, Massachusetts Institute of Technology, 77 Massachusetts Ave, Cambridge, MA 02139, USA}
  \email{kmasui@mit.edu}
\author[0000-0002-2551-7554]{Daniele Michilli}
  \affiliation{Laboratoire d'Astrophysique de Marseille, Aix-Marseille Univ., CNRS, CNES, Marseille, France}
  \email{danielemichilli@gmail.com}
\author[0000-0002-0940-6563]{Mason Ng}
  \affiliation{Department of Physics, McGill University, 3600 rue University, Montr\'eal, QC H3A 2T8, Canada}
  \affiliation{Trottier Space Institute at McGill University, 3550 rue University, Montr\'eal, QC H3A 2A7, Canada}
  \email{mason.ng@mcgill.ca}
\author[0000-0003-0510-0740]{Kenzie Nimmo}
  \affiliation{MIT Kavli Institute for Astrophysics and Space Research, Massachusetts Institute of Technology, 77 Massachusetts Ave, Cambridge, MA 02139, USA}
  \email{knimmo@mit.edu}
\author[0000-0002-8912-0732]{Aaron~B.~Pearlman}
  \altaffiliation{Banting Fellow, McGill Space Institute~(MSI) Fellow, \\ and FRQNT Postdoctoral Fellow.}
  \affiliation{Department of Physics, McGill University, 3600 rue University, Montr\'eal, QC H3A 2T8, Canada}
  \affiliation{Trottier Space Institute at McGill University, 3550 rue University, Montr\'eal, QC H3A 2A7, Canada}
  \email{aaron.b.pearlman@physics.mcgill.ca}
\author[0000-0003-2155-9578]{Ue-Li Pen}
  \affiliation{Institute of Astronomy and Astrophysics, Academia Sinica, Astronomy-Mathematics Building, No. 1, Sec. 4, Roosevelt Road, Taipei 10617, Taiwan}
  \affiliation{Canadian Institute for Theoretical Astrophysics, 60 St.~George Street, Toronto, ON M5S 3H8, Canada}
  \affiliation{Canadian Institute for Advanced Research, 180 Dundas St West, Toronto, ON M5G 1Z8, Canada}
  \affiliation{Dunlap Institute for Astronomy \& Astrophysics, University of Toronto, 50 St.~George Street, Toronto, ON M5S 3H4, Canada}
  \affiliation{Perimeter Institute for Theoretical Physics, 31 Caroline Street N, Waterloo, ON N25 2YL, Canada}
  \email{pen@cita.utoronto.ca}
\author[0000-0002-4795-697X]{Ziggy Pleunis}
  \affiliation{Anton Pannekoek Institute for Astronomy, University of Amsterdam, Science Park 904, 1098 XH Amsterdam, The Netherlands}
  \affiliation{ASTRON, Netherlands Institute for Radio Astronomy, Oude Hoogeveensedijk 4, 7991 PD Dwingeloo, The Netherlands}
  \email{z.pleunis@uva.nl}
\author[0000-0001-7694-6650]{Masoud Rafiei-Ravandi}
  \affiliation{Department of Physics, McGill University, 3600 rue University, Montr\'eal, QC H3A 2T8, Canada}
  \email{masoudrafieiravandi@gmail.com}
\author[0000-0002-4623-5329]{Mawson Sammons}
  \affiliation{Department of Physics, McGill University, 3600 rue University, Montr\'eal, QC H3A 2T8, Canada}
  \affiliation{Trottier Space Institute at McGill University, 3550 rue University, Montr\'eal, QC H3A 2A7, Canada}
  \email{mawson.sammons@mcgill.ca}
\author[0000-0003-3154-3676]{Ketan R. Sand}
  \affiliation{Department of Physics, McGill University, 3600 rue University, Montr\'eal, QC H3A 2T8, Canada}
  \affiliation{Trottier Space Institute at McGill University, 3550 rue University, Montr\'eal, QC H3A 2A7, Canada}
  \email{ketan.sand@mail.mcgill.ca}
\author[0000-0002-7374-7119]{Paul Scholz}
  \affiliation{Department of Physics and Astronomy, York University, 4700 Keele Street, Toronto, ON MJ3 1P3, Canada}
  \affiliation{Dunlap Institute for Astronomy \& Astrophysics, University of Toronto, 50 St.~George Street, Toronto, ON M5S 3H4, Canada}
  \email{pscholz@yorku.ca}
\author[0000-0002-2088-3125]{Kendrick Smith}
  \affiliation{Perimeter Institute of Theoretical Physics, 31 Caroline Street North, Waterloo, ON N2L 2Y5, Canada}
  \email{kmsmith@perimeterinstitute.ca}
\author[0000-0001-9784-8670]{Ingrid Stairs}
  \affiliation{Department of Physics and Astronomy, University of British Columbia, 6224 Agricultural Road, Vancouver, BC V6T 1Z1 Canada}
  \email{stairs@astro.ubc.ca}

\newcommand{\allacks}{
M.B is a McWilliams fellow and an International Astronomical Union Gruber fellow. M.B. also receives support from the McWilliams seed grant.
A.P.C. is a Vanier Canada Graduate Scholar. 
A.M.C. is a Banting Postdoctoral Researcher.
E.F. is supported by NSF grant AST-2407399.
The AstroFlash research group at McGill University, University of Amsterdam, ASTRON, and JIVE is supported by: a Canada Excellence Research Chair in Transient Astrophysics (CERC-2022-00009); the European Research Council (ERC) under the European Union’s Horizon 2020 research and innovation programme (`EuroFlash'; Grant agreement No. 101098079); and an NWO-Vici grant (`AstroFlash'; VI.C.192.045).
V.M.K. holds the Lorne Trottier Chair in Astrophysics \& Cosmology, a Distinguished James McGill Professorship, and receives support from an NSERC Discovery grant (RGPIN 228738-13).
C. L. acknowledges support from the Miller Institute for Basic Research at UC Berkeley.
K.W.M. holds the Adam J. Burgasser Chair in Astrophysics.
D.M. acknowledges support from the French government under the France 2030 investment plan, as part of the Initiative d'Excellence d'Aix-Marseille Universit\'e -- A*MIDEX (AMX-23-CEI-088 AMX-21-IET-016).
M.N. is a Fonds de Recherche du Québec – Nature et Technologies (FRQNT) postdoctoral fellow.
K.N. is an MIT Kavli Fellow.
A.P. is funded by the NSERC Canada Graduate Scholarships -- Doctoral program.
A.B.P. is a Banting Fellow, a McGill Space Institute~(MSI) Fellow, and a Fonds de Recherche du Quebec -- Nature et Technologies~(FRQNT) postdoctoral fellow.
U.P. is supported by the Natural Sciences and Engineering Research Council of Canada (NSERC) [funding reference number RGPIN-2019-06770, ALLRP 586559-23, RGPIN-2025-06396], Canadian Institute for Advanced Research (CIFAR), AMD AI Quantum Astro.
Z.P. is supported by an NWO Veni fellowship (VI.Veni.222.295).
M.W.S. acknowledges support from the Trottier Space Institute Fellowship program.
K.R.S is supported by FRQNT doctoral research award.
P.S. acknowledges the support of an NSERC Discovery Grant (RGPIN-2024-06266).
K.S. is supported by the NSF Graduate Research Fellowship Program.
FRB research at UBC is supported by an NSERC Discovery Grant and by the Canadian Institute for Advanced Research.
}

\begin{abstract}
Among the thousands of observed fast radio bursts (FRBs), a few sources exhibit exceptionally high burst activity observable by many telescopes across a broad range of radio frequencies.
Almost all of these highly active repeaters have been discovered by CHIME/FRB, due to its daily observations of the entire Northern sky as a transit radio telescope.
\frb{} is a source discovered and reported by CHIME/FRB to the community in January 2024;
given its low declination, even the detection of a few bursts hints at a high burst rate.
Following the community announcement of this source as a potentially active repeater, it was extensively followed up by other observatories and has emerged as one of the most prolific FRB repeaters ever observed.
This paper presents the five bursts CHIME/FRB observed from \frb{}, with channelized raw voltage data saved for two bursts.
We do not observe changes in the DM of the source greater than $\sim$1.3~pc~cm$^{-3}$ in our observations over nearly a year baseline.
We find an RM of $\sim+320$~rad~m$^{-2}$.
We do not find evidence for scattering at the level of $<0.3$~ms in the bursts, and we find no evidence for astrophysical scintillation.
In our observations of \frb{}, we see a burst rate $\sim$49x higher than the median burst rate of apparent non-repeaters also discovered by CHIME/FRB.
Each discovery of highly active FRBs 
provides a valuable opportunity to investigate whether there is a fundamental difference between repeating and apparently non-repeating sources.
\end{abstract}


\section{Introduction}
\label{sec:intro}

Fast radio bursts (FRBs) are one of the most fascinating astrophysical phenomena observed in recent decades.
These bursts have short time durations \citep[typically $\sim$$\mu$s--ms, e.g.,][]{petroff+2022_frbreview}, high brightnesses \citep[$\sim$Jy; ][]{petroff+2022_frbreview} and originate from extragalactic, even cosmological distances \citep{gordon+2023_hosts, connor+2024_igm}.
A subset of FRBs are observed to repeat \citep[e.g.,][]{spitler+2016_frb121102, RN3}, enabling targeted follow-up observational campaigns from various observatories.
Some of the most extensively studied FRB sources are
sources that exhibited unusually active radio burst activity over a time interval of $\sim$weeks, observable across a broad range of radio frequencies.
Such highly active sources, sometimes referred to as ``hyperactive repeaters'', include FRB~20201124A \citep{lanman+2021_R67, kirsten+2024_r67} and FRB~20220912A \citep{r117_atel}, the latter which reached observed burst rates of up to 390~hr$^{-1}$ \citep{zhang+2023_r117FAST}.
Both of these sources were discovered by the Canadian Hydrogen Intensity Mapping Experiment/FRB project (CHIME/FRB) and announced to the community via Astronomer's Telegram (ATel\footnote{\url{https://astronomerstelegram.org/}}) notices.

Announcing the discovery of FRBs during their periods of heightened burst activity has enabled the community to focus their observations on these targets.
For example, the highly active nature of FRB~20201124A and FRB~20220912A were unique testbeds for probing the maximum energetics of FRBs \citep{kirsten+2024_r67, Ould-Boukattine+2024_energetics}.
Alerting the community of heightened source activity from a repeater also allows for quicker VLBI localizations \citep[e.g.,][]{nimmo+2022_R67_evnloc}, probes of RM and DM evolution \citep[e.g.,][]{zhang+2023_r117FAST}, and periodicity searches \citep[e.g.,][]{du+2024_periodicitysearch}.

Here we report the discovery of \frb{} following our ATel announcement in January 2024 during \frb{}'s outburst.
\frb{} is one of the most recent examples of such an active FRB source, fruitfully followed up by many radio telescopes around the world.
Soon after the ATel announcement \citep{r147_atel_chime}, the active nature of this source was observed by observations with the Parkes/Murriyang telescope \citep{Uttarkar+2024ATel_r147};
small European radio dishes including the 25-m Westerbork RT1 telescope, the 25-m Stockert telescope, the 32-m Torun telescope, the 25-m Onsala O8 telescope, and the 25-m Dwingeloo telescope \citep{Ould-Boukattine+2024ATel_r147_first, Ould-Boukattine+2024ATel_r147};
the Five-hundred-meter Aperture Spherical radio Telescope \citep[FAST;][]{Zhang+2024ATel_r147};
the Northern Cross radio telescope \citep{2024ATel16434....1P};
MeerKAT \citep{tian+2024_r147meerkat};
the upgraded Giant Metrewave Radio Telescope \citep{2024arXiv240509749P, 2024ApJ...977..177K};
European Very Long Baseline Interferometry (VLBI) Network (EVN) dishes as part of the PRECISE Project \citep{r147_atel_evn};
the Nançay Radio Telescope \citep{2024ATel16597....1H};
the Allen Telescope Array \citep{2024ATel16599....1J};
the 100-m Effelsberg telescope \citep{Limaye+2024ATel_r147};
and the Robert C. Byrd Green Bank Telescope \citep{xie+2024_r147_FAST}.
These community observations took place between mid-January 2024 and mid-May 2024 and spanned a broad range of radio frequencies, from 300~MHz up to 6~GHz.

The prompt follow-up observations resulted in a variety of fast-paced discoveries.
Observations with the MeerKAT telescope enabled the localization of \frb{} to a dwarf galaxy at $z = 0.1306 \pm 0.0002$ behind a foreground galaxy cluster \citep{2024ATel16426....1O, 2024ATel16613....1B, tian+2024_r147meerkat, chen+2025ApJ_r147host}.
EVN/PRECISE observations have obtained a preliminary localization of \frb{} to $\sim$milliarcseconds precision \citep{r147_atel_evn}.
Using simultaneous observations with Effelsberg, the Thai National Radio Telescope, the Astropeiler Stockert, and the X-ray satellite \textit{XMM-Newton},
X-ray-to-radio fluence ratios were constrained to $\eta_{x/r} < 2.4 \times 10^6$, assuming a cutoff power law physically motivated by the Galactic magnetar SGR~1935$+$2154.
This constraint demonstrates that we cannot rule out the possibility of \frb{} having a shared emission mechanism with SGR~1935$+$2154 \citep{Eppel+2025_r147_xrayradio}.
VLBA observations identified a possible persistent radio source (PRS) associated with \frb{} \citep{bruni+2024_r147prs}.
Follow-up MeerKAT and VLA observations conducted 26 days after the initial burst identified a possible compact flaring radio source, with spectrotemporal properties distinct from those of PRSs \citep{2025arXiv250114247Z}.
Gamma-ray emission from the \textit{Fermi} instrument coincident with \frb{} was also reported \citep{2024arXiv241106996X}, though the significance of this claim has been disputed by the \textit{Fermi}-LAT Collaboration \citep{2024ATel16602....1P}.

In this paper, we provide details on the original detection of \frb{} \citep{r147_atel_chime}, one of the most prolific FRB sources known thus far.
We describe our observations with CHIME/FRB in Section~\ref{sec:observations}, the repetition rate in Section~\ref{sec:repetition}, and burst properties in Section~\ref{sec:burst_props}.
In Section~\ref{sec:disc_conc}, we conclude with a discussion of the larger context of the source's repetition rate and burst properties.

\section{Observations}
\label{sec:observations}

\begin{figure*}
    \centering
    \subfloat{{\includegraphics[width=0.47\textwidth]{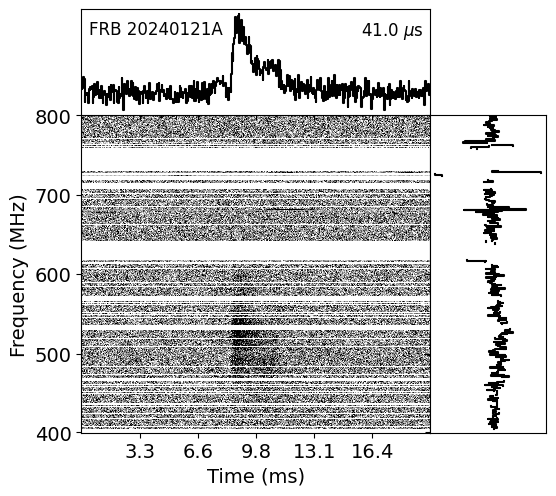} }}%
    \quad
    \subfloat{{\includegraphics[width=0.47\textwidth]{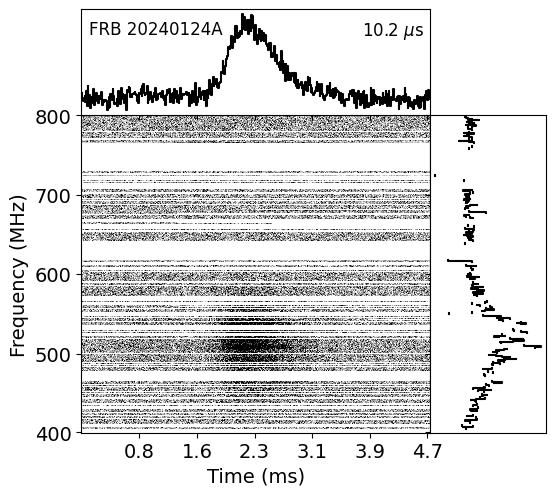} }}%
    \caption{Dynamic spectra of the two bright bursts from \frb{}, both with baseband data saved, plotted at a frequency resolution of 195.3125~kHz.
    RFI-contaminated frequency channels are masked; in some channels, there are also missing data due to networking issues.
    Each burst was de-dispersed to its structure-maximizing DM of $\approx$527.7~pc~cm$^{-3}$ (Table~\ref{tab:burst_props}).
    Displayed with each dynamic spectrum subplot (main subpanels) are the TNS names (top left label), the time resolutions of the plotting (top right label), the time-integrated spectra (right subpanels), and the frequency profiles (top subpanels).
    }
    \label{fig:waterfalls}
\end{figure*}

The CHIME telescope is located at the Dominion Radio Astronomical Observatory (DRAO) near Penticton, British Columbia, Canada.
As a transit telescope, it surveys the Northern sky (declination $> -11^\circ$) roughly every day at 400$-$800~MHz \citep{chime_sys_overview, chimefrb_sys_overview}.
The CHIME/FRB system currently has by far the highest discovery rate of new FRB sources of all the presently operating radio observatories \citep{catalog1}.
All candidate events with a detection signal-to-noise (S/N) greater than 8 have intensity (Stokes $I$) data saved, at 0.983~ms time resolution.
For bursts with S/N $> 12$ and bursts from candidate repeaters with S/N $> 10$, channelized raw voltage data (``baseband'') are saved with a time resolution of 2.56~$\mu$s \citep{michilli+2021_baseband_pipeline}.

\frb{} was first detected on 2024 January 14 with S/N~$\sim$~8 and a DM of $\sim$528~pc~cm$^{-3}$.
Two more bright bursts were observed 4 days apart later that month (Figure~\ref{fig:waterfalls});
these two events triggered the CHIME/FRB baseband system and were streamed through the CHIME/FRB VOEvents service \citep{abbott+2025_voevents}.

\section{Repetition Rate}
\label{sec:repetition}

While CHIME/FRB has observed a number of sources with higher apparent burst rates \citep[e.g.,][]{lanman+2021_R67, r117_atel, shah+2025_r155, r157_atel}, what was remarkable about \frb{} was its low (4.3$^\circ$) declination.
At such a low declination, CHIME's daily exposure to \frb{} is $\sim$5.5~min per day, compared to a ``typical'' daily exposure of $\sim$15~min near zenith.
We consider exposure to be when the position is within the full-width half-maximum (FWHM) of our formed beams at 600~MHz, estimated using the CHIME beam model,\footnote{\url{https://chime-frb-open-data.github.io/beam-model/}} when the system is completely on and sensitive.
More details can be found in \citet{catalog1}

The detection of two bright bursts within four days of each other at such a low declination hinted at a high burst rate.
For those two bursts, baseband data were saved, which allowed us to obtain a localization of \frb{} to $\sim$arcminute precision \citep{michilli+2021_baseband_pipeline}.
This active source, along with its localization, was promptly announced to enable rapid community follow-up observations \citep{r147_atel_chime}. 

From 2018 August 28 through 2025 February 14, the total CHIME/FRB exposure to \frb{} at its best-known position \citep{r147_atel_evn} is $\sim$104~hr.
Three bursts (described below, but for which no baseband data were saved) were detected in the above time interval within the FWHM of our formed beams at 600~MHz.
(The two bursts for which baseband data were saved were detected outside the FWHM, and thus are not considered in our exposure/rate calculations.)
Using the methodology described by \citet{josephy+2019_R1}, we estimate a 95\% fluence threshold of of 15.65~Jy~ms.
We thus obtain an overall burst rate of 0.03~hr$^{-1}$ above a fluence threshold of 15.65~Jy~ms.

We can also compare the CHIME/FRB-observed burst rate of \frb{} to those of the apparent non-repeaters in the CHIME/FRB Catalog~2 sample (CHIME/FRB Collaboration et al. in prep.), assuming a cumulative fluence index of $\alpha = -1.5$ to scale to the same fluence threshold for all bursts.
We find \frb{} has a CHIME/FRB-observed burst rate $\sim$49x higher than the median burst rate of apparent non-repeaters in Catalog~2,
and $\sim$12x higher than the median burst rate of apparent non-repeaters at a similar declination ($4^\circ \leq \rm{Dec} \leq 5^\circ$).

The ``hyperactive'' behavior of \frb{} was soon confirmed by other radio telescopes at a variety of observing frequencies \citep[e.g.,][]{Uttarkar+2024ATel_r147, Zhang+2024ATel_r147, tian+2024ATel_r147, Ould-Boukattine+2024ATel_r147, Limaye+2024ATel_r147}.
One 1.38-hr observing session with the Robert C. Byrd Green Bank Telescope resulted in 359 bursts \citep{xie+2024_r147_FAST}.
Additionally, a 30-min observation with FAST revealed a burst rate of $\sim$500 hr$^{-1}$ above a fluence threshold of 0.015~Jy~ms in the FAST band of 1.0 to 1.5 GHz \citep{zhang+2024ATel2_r147}, the highest known burst rate quoted for any FRB to-date.

We note that \frb{}, originating from a redshift of $z \sim 0.13$, is more distant than the other extremely active sources FRB~20201124A \citep[$z = 0.0979 \pm 0.0001$;][]{2021ATel14516....1K, fong+2021_r67} and FRB~20220912A \citep[$z = 0.0771 \pm 0.0001$;][]{ravi+2023_r117}, establishing that the high burst rates of these ``hyperactive'' repeaters are not due solely to an excess of low-luminosity bursts.

\section{Burst properties from CHIME/FRB}
\label{sec:burst_props}

In order to infer burst properties of \frb{} from our CHIME/FRB observations, we use a least-squares optimization fitting framework called \texttt{fitburst} \citep{fonseca+2023_fitburst}.
This routine can be performed on the dynamic spectra of both intensity and baseband data \citep[e.g.,][]{sand+2023_R3, sand+2025_basecat1morph,curtin+2024_RNmorph} and can estimate the following burst properties:
dispersion measure (DM), time of arrival (ToA), signal amplitude, pulse width, scatter-broadening timescale ($\tau$), spectral index, and spectral running.
If a burst appears to consist of multiple components, \texttt{fitburst} will estimate the signal amplitude, pulse width\footnote{This is the 1-$\sigma$, one-sided width.}, ToA, spectral index, and spectral running separately for each subpulse.
The scatter-broadening timescale profile is assumed to depend on frequency as $\tau \propto \nu^{-4}$.
Additionally, scattering timescales determined solely using intensity data can be unreliable, as downward drifting or unresolved components can easily be interpreted as scattering at low time resolutions. Additionally, given the measurement is in conflict with the limits using the higher resolution baseband data, this strongly suggests that this is not scattering and instead unresolved downward drifting. 
Hence, we caution over-interpretation of the single scattering measurement made for this source using intensity data. 

Fluxes and fluences are derived separately from burst fitting with \texttt{fitburst}, as they depend strongly on the CHIME beam model.
For intensity data, reported fluxes and fluences are lower limits because of the large localization uncertainty of the burst within the CHIME synthesized beam.
However, if an accurate position for a known source exists, then the synthesized beam attenuation can be corrected \citep{andersen+2023_chimefrb_flux};
we apply this technique in this work for \frb{}.
Fluxes and fluences for baseband data are also corrected for the beam response \citep{basecat1}.
Thus, all flux and fluence values derived for \frb{} are considered measurements (rather than lower limits for the intensity data).
The resulting measured burst properties are summarized in Table~\ref{tab:burst_props}.
We note that the DM values obtained from intensity data likely have underestimated uncertainties due to unresolved burst sub-structure.

\begin{deluxetable*}{lccccccc}
\tabletypesize{\footnotesize} 
\tablewidth{0pt}
\tablecaption{Summary of detections of \frb{} from CHIME/FRB. \label{tab:burst_props}}
\tablehead{
    \colhead{TNS name} & \colhead{ToA} & \colhead{S/N} & \colhead{DM} & \colhead{Width} & \colhead{$\tau$} & \colhead{Peak Flux} & \colhead{Fluence} \\
    \colhead{} & \colhead{} & \colhead{} & \colhead{(pc cm$^{-3}$)} & \colhead{(ms)} & \colhead{(ms)} & \colhead{(Jy)} & \colhead{(Jy ms)}
}
\startdata
    FRB~20240114A & 2024-01-14 21:50:39 & 8.17 & $528.98 \pm 0.15$ & $2.1 \pm 0.1$ & $< 0.42$ & $1.2 \pm 0.7$ & $7 \pm 2$ \\ 
    FRB~20240121A\tablenotemark{*} & 2024-01-21 21:30:40 & 18.29 & $527.68 \pm 0.02$ & [$0.35 \pm 0.02$, $0.83 \pm 0.08$]\tablenotemark{a} & $<0.35$ & $91 \pm 11$ & $146 \pm 19$ \\
    FRB~20240124A\tablenotemark{*} & 2024-01-24 21:20:11 & 32.37 & $527.69 \pm 0.03$ & $0.330 \pm 0.006$ & $<0.330$ & $1121 \pm 115$ & $1014 \pm 108$ \\
    FRB~20240224B & 2024-02-24 19:11:24 & 9.16 & $528.33 \pm 0.17$ & $3.1 \pm 0.2$ & $< 0.61$ & $2.5 \pm 1.2$ & $10 \pm 2$ \\ 
    FRB~20250101A & 2025-01-01 22:42:08 & 8.72 & $528.75 \pm 0.11$ & [$1.0 \pm 0.2$, $1.6 \pm 0.7$]\tablenotemark{a} & $0.51 \pm 0.28$\tablenotemark{b} & $1.3 \pm 0.6$ & $2 \pm 1$ 
\enddata
\tablecomments{The ToA timestamps are referenced at 400 MHz (topocentric at CHIME near Penticton, Canada) and calculated using the DM indicated in the fourth column. S/N is as reported by the real-time detection pipeline. All uncertainties are quoted at the 1-$\sigma$ level.
The scattering timescales ($\tau$) are referenced to 600~MHz, assuming a scattering index of $-4$.
In the case of an upper limit on the scattering timescale, they are reported as the 1-$\sigma$, one-sided width of the narrowest subburst. Fluences are averaged over the entire 400--800~MHz observing band of CHIME.}
\tablenotemark{*}{Bursts for which baseband data were saved. For these bursts, burst parameters were obtained by using \texttt{fitburst} on the baseband data.}

\tablenotemark{a}{The two reported widths correspond to the those of two distinct sub-bursts identified in the burst envelope.}

\tablenotemark{b}{We caution that this source shows downward drifting, and hence a measurement of scattering at a lower time resolution is likely unresolved downward drifting. }
\end{deluxetable*}

\subsection{Scintillation}
\label{subsec:scintillation}

Scintillation, observable as a frequency-dependent modulation in brightness, can provide insights into the local environment of radio bursts and the intervening medium.
The scintillation bandwidth follows the relation $\Delta \nu_\mathrm{d} \propto \nu^{-\alpha}$, with $\alpha \sim 4$ \citep{lorimerkramer2012_handbook}.
Using the NE2001 Galactic electron density model \citep{ne2001}, the expected Galactic scintillation bandwidth at 1 GHz is $\Delta \nu_\mathrm{d} = 0.59$~MHz.
Assuming $\alpha = 4$, the predicted value at $600$~MHz is $\Delta \nu_\mathrm{d} = 0.08$~MHz.
In the baseband data, the frequency resolution is $0.39$~MHz, but upchannelization can allow for finer-scale scintillation analysis at the cost of time resolution \citep[e.g.,][]{schoen2021scintillation, nimmo+2024_scintillation}.

We use the scintillation pipeline detailed by \citet{nimmo+2024_scintillation} to search for scintillation scales in the autocorrelation functions of the burst spectra.
We examine the two sets of saved baseband data for scintillation signatures at up-channelization factors of 32, 128, and 512, corresponding to frequency resolutions of 0.01, 0.003, and 0.0008~MHz, respectively. 
The duration of the two baseband bursts are $\sim 1$~ ms, which implies an intrinsic Fourier-limited frequency resolution limit of $\sim 0.001$~MHz.
Our highest up-channelization factor exceeds this limit.

We find no convincing evidence for astrophysical scintillation, either extragalactic or Galactic.
J.~Huang et al. in prep. report an L-band burst from \frb{} observed by the Nançay Radio Telescope that shows scintillation barely unresolved in the 4~MHz observing channels, but consistent with NE2001 expectations.
The non-detection of scintillation from the pipeline used on CHIME/FRB data could be due to systematics, self-noise, or decoherence.
Quantifying the completeness of the CHIME/FRB scintillation pipeline to such effects is active work that is beyond the scope of this paper.

\subsection{Polarization}
\label{subsec:polarization}

\begin{figure}
    \centering
    \includegraphics[width=\columnwidth]{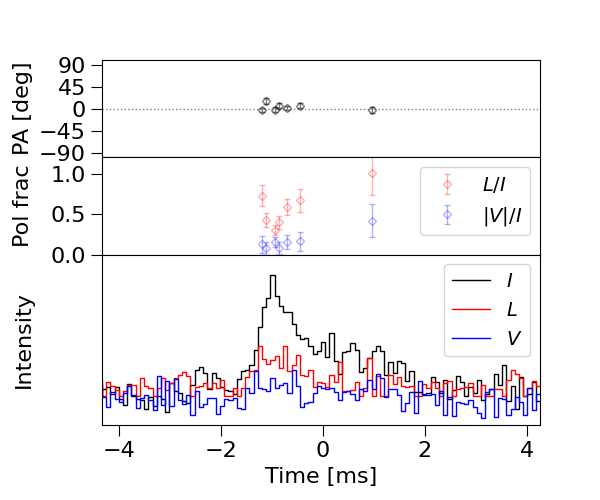}
    \caption{The polarization profile for FRB~20240121A.
    The top panel shows the polarization position angle.
    The middle panel shows the binned linear (red) and circular (blue) polarization component fractions.
    The bottom panel shows the frequency-averaged burst profile in arbitrary units for the total intensity (black), de-biased linear polarization, and circular polarization.
    An RM of $+324.4$~rad~m$^{-2}$ was used to de-rotate the burst.
    }
    \label{fig:polprof}
\end{figure}

We also explored the polarimetric properties of \frb{} for our two bursts with saved baseband data (TNS names FRB~20240121A and FRB~20240124A, respectively).
The pulse, linear and circular polarization fraction, and PA profiles of the brighter of the two bursts, FRB~20240121A, are shown in Figure~\ref{fig:polprof}.
We see a significant level of instrumental leakage in both bursts which manifests as a peak in the Faraday dispersion function (FDF) centered at $\mathrm{RM} = 0~\mathrm{rad}~\mathrm{m}^{-2}$.
Following \cite{pandhi+2024_basecat_pol}, we mask out the peak at $0~\mathrm{rad}~\mathrm{m}^{-2}$ and search for a secondary peak in the FDF exceeding a linearly polarized S/N of 6.
Using the RM-synthesis algorithm \citep{burn_1966, brentjens_deBruyn_2005_RM_synth} as implemented by \citet{mckinven+2021_polarization_pipeline}, we obtain an improved polarized fit for both bursts with a secondary FDF peak at $|\mathrm{RM}| > 0~\mathrm{rad}~\mathrm{m}^{-2}$. 
For FRB~20240121A, we obtain an RM of $+324.4 \pm 0.3$~rad m$^{-2}$ and a linear polarization fraction $L/I = 0.45 \pm 0.03$.
For FRB~20240124A, we find a peak in the FDF at $\mathrm{RM} = +320~\mathrm{rad}~\mathrm{m}^{-2}$ with $L/I = 0.44 \pm 0.01$.
However, the FDF of this burst is Faraday complex, which could lead an underestimation of its RM uncertainty which, for Faraday simple peaks, is quantified as $\mathrm{FWHM}/(2 \ \mathrm{S/N})$, where the FWHM and S/N are the full-width and half-maximum and the S/N of the peak polarized intensity in the FDF.
Thus, we apply a more conservative uncertainty for FRB~20240124A equal to width of the theoretical RM spread function of CHIME/FRB \citep[$9~\mathrm{rad}~\mathrm{m}^{-2}$;][]{ng+2019_pulsar} and determine $\mathrm{RM} = 320 \pm 9~\mathrm{rad}~\mathrm{m}^{-2}$. 
We note that due to the instrumental leakage present in both bursts, it is possible that our L/I measurements are underestimated.
The circular polarization fraction in both bursts is consistent with instrumental effects.

The Galactic foreground RM expectation in the direction of \frb{} is $-15 \pm 10$~rad m$^{-2}$ \citep{hutschenreuter+2022_MW_RM_map}.
We do not find any RM variation between the two polarized bursts in our sample.
Our RM results are consistent with those reported from other observations of \frb{} at higher frequencies \citep[e.g.,][]{tian+2024_r147meerkat, xie+2024_r147_FAST}.
The RM amplitude is also fairly large compared to the median $|\mathrm{RM}|$ from non-repeating \citep{pandhi+2024_basecat_pol} and repeating \citep{ng+2025_rn3pol} sources.

\section{Discussion \& Conclusion}
\label{sec:disc_conc}

Here, we present the initial detection of the hyper-active repeating FRB, FRB 20241104A, with CHIME/FRB. While we only detect five bursts from this source over a six year period, the source's low declination, and hence short transit time, imply a high burst rate. Indeed, we determine a rate of 0.03 bursts hr$^{-1}$ above a fluence threshold of 15.65 Jy ms for this source in CHIME's 400 to 800 MHz band. Assuming the rate scales as a power-law with an index of $-1.5$, this corresponds to a rate of 0.16 bursts hr$^{-1}$ above a fluence threshold of 5 Jy ms. Compared to the CHIME/FRB repeater rates presented by \citet{RN3}, this source ranks in the top 15\% of active repeaters detected by CHIME. Comparing its rate at CHIME frequencies to those at other bands is non-trivial, as instrument selection effects and sensitivity functions must be carefully accounted for. We therefore do not pursue this comparison further in this work.

The transit telescope nature of CHIME/FRB is a double-edged sword ---
while the observing strategy allowed for the serendipitous detection of \frb{}, which in turn led to rapid and fruitful observations by the larger transient radio community,
the limited exposure and lower sensitivity at lower declinations meant that CHIME/FRB only observed five bursts from \frb{}, of which only two had bright enough S/N to obtain meaningful burst properties.
Thus, in-depth analysis of its DM, scattering timescales, scintillation, and polarization properties were only possible for two out of the five bursts.

Observations with MeerKAT discovered 62 bursts from this source \citep{tian+2024_r147meerkat}. In contrast with our work, they find most of the bursts are 100\% linearly polarized, and that some bursts show circular polarization fractions up to 20\%. This is similar to the results of \citet{xie+2024_r147_FAST} who find that 72\% of the 297 bursts they detected with Robert C. Byrd Green Bank Telescope had linear polarization fractions greater than 90\%. While these results are in slight contrast with the relatively lower linear polarization fraction found in our bursts, our sample size is particularly small. 

\citet{tian+2024_r147meerkat} also do not find evidence for scattering, with a limit of $<0.4 \pm 0.2$ ms at 1 GHz or $<4$ms at 600 MHz assuming a $\nu^{-4}$ scaling. This is an order of magnitude less conservative than our results, consistent with expectations as higher-frequency observations have reduced sensitivity to scattering tails. 

One additional benefit of CHIME's observations is that it can monitor for renewed burst activity ``for free'' over extended temporal periods.
Indeed, the last CHIME/FRB detection of a burst from \frb{} in early Jan 2025, after over 10 months of non-detections, coincided with observations of renewed burst activity by the HyperFlash project \citep{Ould-Boukattine+2025ATel2_r147}.

\frb{} raises the question of how many prolific repeating FRB sources at low declinations may have been missed in spite of CHIME's enormous, yet finite, field-of-view.
Such sources definitely exist --- MeerKAT observed FRB~20240619D, a source with an average burst rate of $>$70/hr at L-band above a $\sim$1~Jy~ms fluence threshold \citep{Tian+2024ATel}.
Yet without a transit telescope similar to CHIME in the Southern hemisphere, it is probable many extremely active FRB sources have gone undetected.

\frb{} has demonstrated the power of quickly announcing the discovery of a new FRB source with heightened burst activity to the broader FRB community.
No one radio telescope alone could have discovered \frb{} and followed it up as extensively as \frb{} has been studied.
We anticipate many new insights into the origins of FRBs to be enabled by the extensive observations of \frb{}.

\begin{acknowledgments}

We acknowledge that CHIME is located on the traditional, ancestral, and unceded territory of the Syilx/Okanagan people. We are grateful to the staff of the Dominion Radio Astrophysical Observatory, which is operated by the National Research Council of Canada. CHIME operations are funded by a grant from the NSERC Alliance Program and by support from McGill University, University of British Columbia, and University of Toronto. CHIME was funded by a grant from the Canada Foundation for Innovation (CFI) 2012 Leading Edge Fund (Project 31170) and by contributions from the provinces of British Columbia, Québec and Ontario. The CHIME/FRB Project was funded by a grant from the CFI 2015 Innovation Fund (Project 33213) and by contributions from the provinces of British Columbia and Québec, and by the Dunlap Institute for Astronomy and Astrophysics at the University of Toronto. Additional support was provided by the Canadian Institute for Advanced Research (CIFAR), the Trottier Space Institute at McGill University, and the University of British Columbia. The CHIME/FRB baseband recording system is funded in part by a CFI John R. Evans Leaders Fund award to IHS.

\allacks

\end{acknowledgments}

\begin{contribution}

K.S. led the analysis and writing of the paper.
A.C., M.F., A.P. contributed to analysis and writing of the paper.
All other authors are listed alphabetically and were instrumental in acquiring the observations or contributed relevant scientific expertise to the project.


\end{contribution}

\facility{CHIME}

\software{
    \texttt{fitburst} \citep{fonseca+2023_fitburst},
    \texttt{matplotlib} \citep{matplotlib},
    \texttt{NE2001p} \citep{ocker_cordes_2024_NE2001p},
    \texttt{numpy} \citep{numpy},
    \texttt{PyGEDM} \citep{pygedm},
    \texttt{RM-CLEAN} \citep{heald+2009_rmclean},
    \texttt{RM-synthesis} \citep{brentjens_deBruyn_2005_RM_synth}, 
    \texttt{RM-tools} \citep{RMTools},
    \texttt{scipy} \citep{scipy}
}

\appendix

\section{Dynamic spectra for intensity-only bursts}

\begin{figure}[h]
    \centering
    \includegraphics[width=0.98 \linewidth]{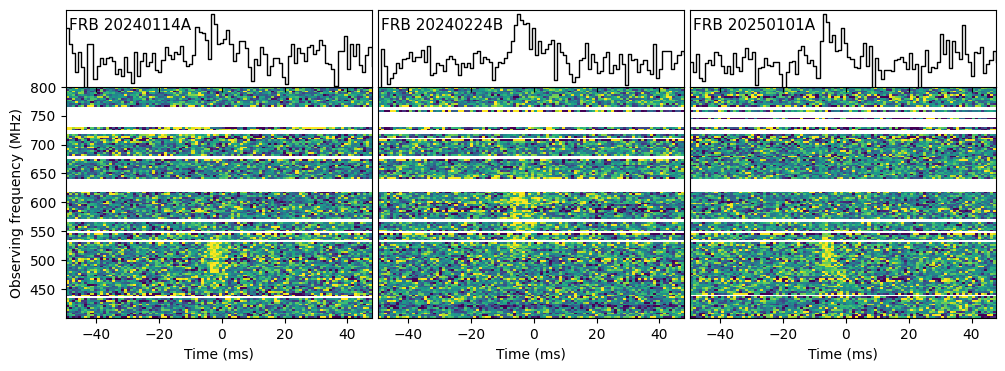}
    \caption{
    Dynamic spectra of the three bursts from FRB~20240114A for which only intensity data were saved.
    Each burst is de-dispersed to the best-fit DM determined by \texttt{fitburst} (in Table~\ref{tab:burst_props}).
    Displayed with each dynamic spectrum subplot (main subpanels) are the TNS names (top left label) and the frequency profiles (top subpanels).
    Data are plotted at 3.125~MHz frequency resolution and 0.983~ms time resolution.
    }
    \label{fig:intensity_waterfalls}
\end{figure}

For three bursts, we do not have baseband data saved.
In Figure~\ref{fig:intensity_waterfalls}, we show the waterfall plots for the bursts with intensity data.

\bibliography{refs}{}
\bibliographystyle{aasjournal}

\end{document}